\documentclass[a4paper,11pt]{article}
\usepackage{acl2010}
\usepackage{times}
\usepackage{url}
\usepackage{latexsym}
\usepackage{graphicx}

\graphicspath{{images/}{./}}

\newcommand{\QANUS}{\textbf{\textsc{Qanus}}}
\newcommand{\QANUSSYS}{\textbf{\textsc{Qa-Sys}}} 
\newcommand{\HG}{\textsc{HG-01}}

\title{QANUS: An Open-source Question-Answering Platform}

\author{Jun-Ping Ng\\
  Department of Computer Science\\
  National University of Singapore\\
  {\tt junping@comp.nus.edu.sg}  \And
  Min-Yen Kan\\
  Department of Computer Science\\
  National University of Singapore\\
  {\tt kanmy@comp.nus.edu.sg}}

\date{}

\begin{document}
\maketitle
\begin{abstract}

In this paper, we motivate the need for a publicly available, generic software framework for question-answering (QA) systems. We present an open-source QA framework \QANUS\ which researchers can leverage on to build new QA systems easily and rapidly. The framework implements much of the code that will otherwise have been repeated across different QA systems. To demonstrate the utility and practicality of the framework, we further present a fully functioning factoid QA system \QANUSSYS\ built on top of \QANUS .
\end{abstract}

\section{Introduction}


There has been much research into question-answering (QA) over the past decades. However the community is still lacking QA systems which are readily available for use. This translates into a high barrier of entry for researchers who are new to the field. The absence of easily accessible systems also means that there is a lack of credible, reproducible baseline systems against which new QA systems can be evaluated.


To address the highlighted limitations, we are releasing an open-source, Java-based, QA framework \QANUS\ (pronounced KAY-NESS). \QANUS\ is a framework on which new QA systems can be easily and rapidly developed. \QANUS\ makes it easy to build new QA systems as only a minimal set of components needs to be implemented on top of the provided framework. To demonstrate the utility and practicality of \QANUS , a reference implementation of a QA system \QANUSSYS\ has also been developed using the framework. \QANUSSYS\ is also made available to the community. When it matures, it can serve as an accessible, reproducible baseline system for evaluations.

To ensure the availability of the system to the community, as well as to maximise the benefits of any derivative projects for everyone, \QANUS\ is released under the Open Software License (OSL) v3.0.

\section{Related Work}

\begin{figure}[htbp]
	\centering
		\includegraphics[width=0.95\linewidth]{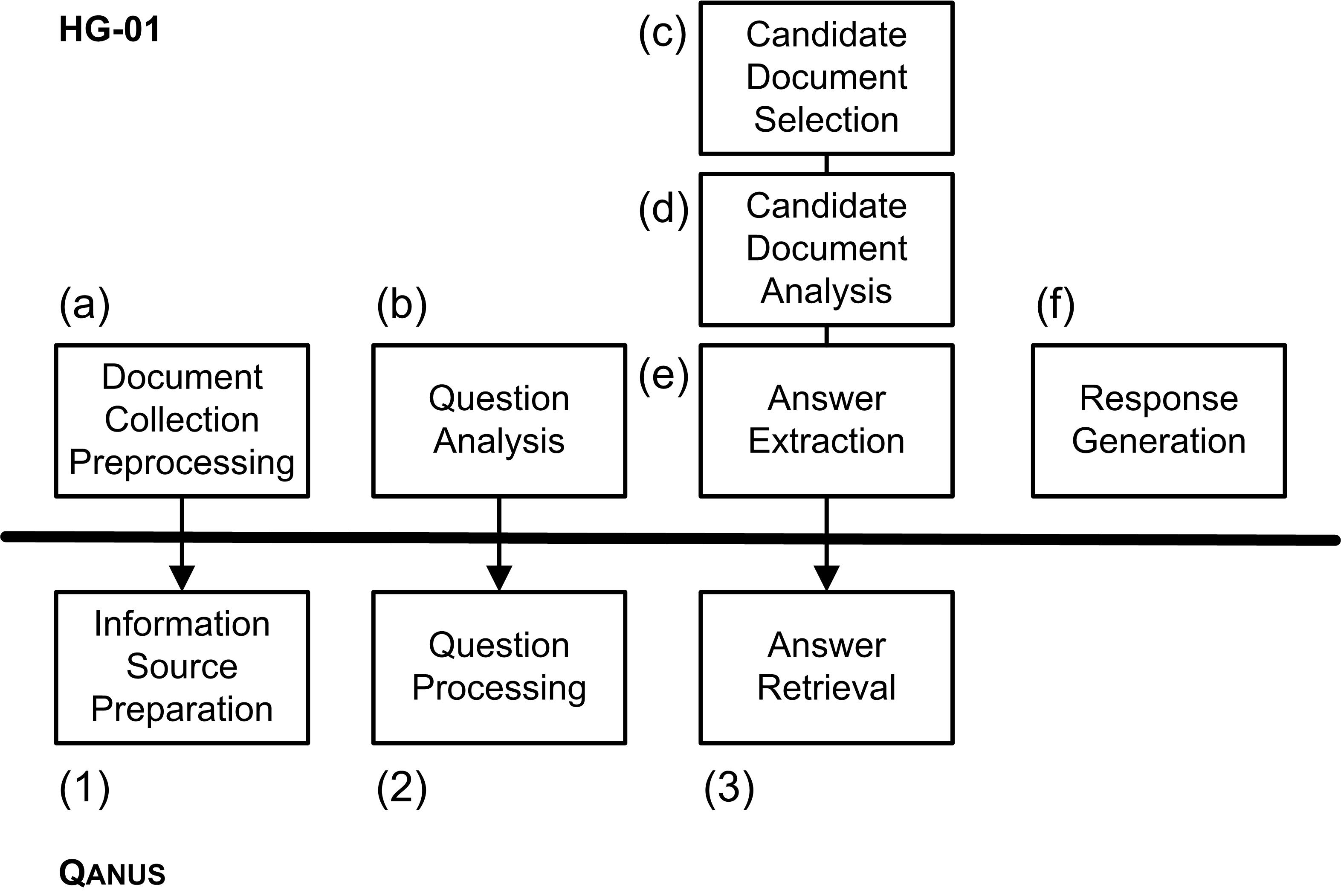}
	\caption{Comparing pipeline stages of \HG\ and \QANUS .}
	\label{fig:hirschman-compare}
\end{figure}

There has been previous efforts in generalising the architecture of QA systems. Hirschman and Gaizauskas~\shortcite{Hirschman2001} for example described a pipelined approach to QA (\HG ), where different stages are combined serially into a QA system. Figure~\ref{fig:hirschman-compare} highlights the different stages in their pipeline vis-a-vis the stages found in \QANUS . The informal correspondence between the various stages of the two pipelines are also shown in the figure.

The architecture of \HG\ is slanted towards QA systems based on current state-of-the-art information retrieval (IR) techniques. These techniques typically involve manipulating the lexical and syntactic form of natural language text and do not attempt to comprehend the semantics expressed by the text. Systems which make use of these techniques \cite{Hickl2007,Chali2007} have been able to perform ahead of their peers in the Text Retrieval Conference (TREC) QA tracks \cite{Dang2007}.

In IR-based systems, answer processing revolves around units of information stored in documents. To reflect the importance of this organisation two separate stages \emph{(c) candidate document selection} and \emph{(d) candidate document analysis} are described in Hirschman's architecture. Further, \emph{(f) answer generation} is included as they considered interactive QA systems which could participate in a dialogue with end-users. 

Not all QA systems are IR-centric however, and interactive QA systems are likely not imminent given the limitations of natural language understanding and generation. \QANUS\ thus generalises stages \textit{(c)}, \textit{(d)} and \textit{(e)} into one to avoid over-committing to any particular architecture or paradigm, and leaves out \textit{(f)}. 

Another important point of comparison is that \QANUS\ is an implemented, functional QA architecture whereas \HG\ serves mainly as a general discussion and introduction to the architecture of QA systems.


Though few in numbers, some QA systems have previously been made available to the community. One such system is \textsc{Aranea}\footnote{Available for download at http://www.umiacs.umd.edu/$\sim$jimmylin/downloads/index.html} \cite{Lin2007}. \textsc{Aranea} is a factoid QA system which seeks to exploit the redundancy of data on the web and has achieved credible performances at past TREC evaluations. \textsc{Aranea} is not designed however as a generic QA platform.
We argue that a framework such as \QANUS\ which is designed from the start with extensibility and flexibility in mind will greatly reduce the effort needed for any such customisation.

\textsc{Qanda} by \textsc{Mitre}\footnote{http://www.openchannelsoftware.org/projects/Qanda} is another QA system which has featured in the TREC QA track. It has a project page on SourceForge. However currently only one module of the system is made available for download. We are at the time of writing unable to verify if there are plans for the release of the rest of the system in the near future.

\section{\QANUS\ Framework}


\begin{figure*}[htbp]
	\centering
		\includegraphics[width=0.90\linewidth]{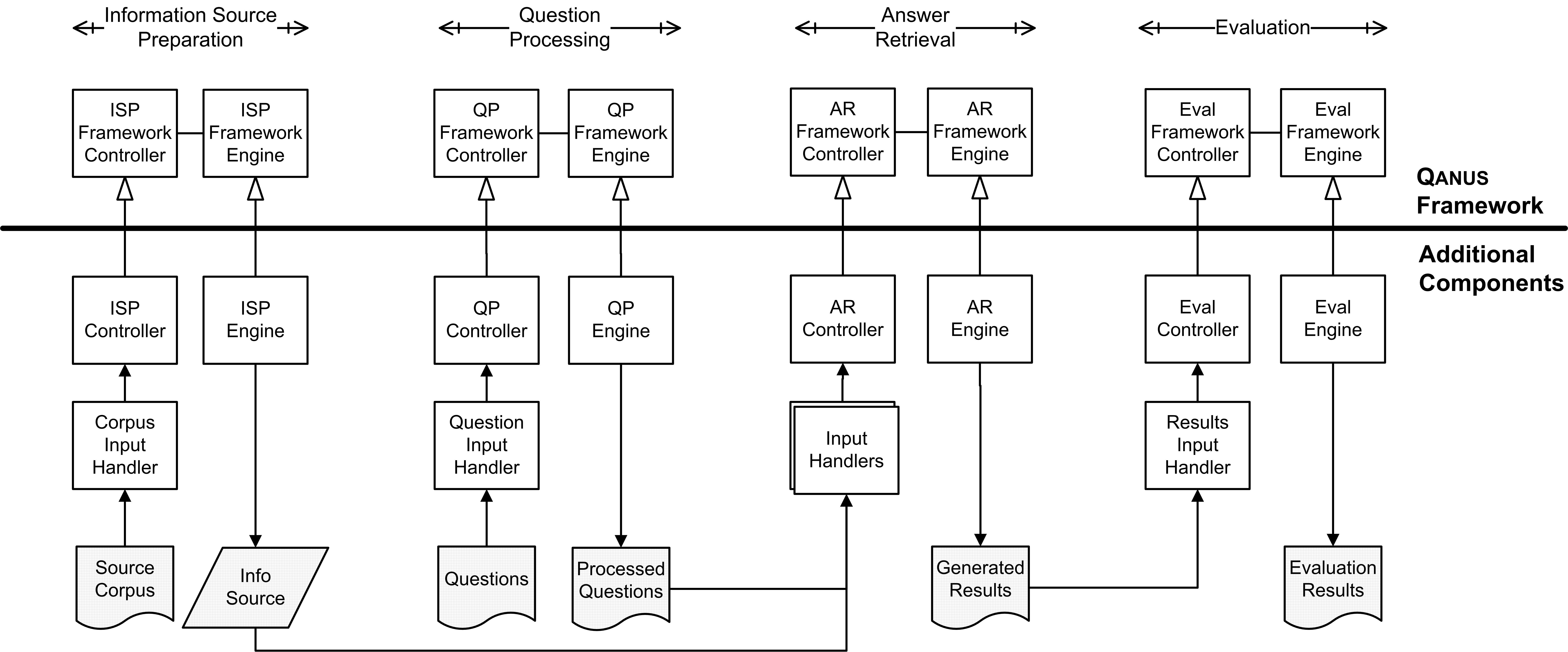}
	\caption{Full QA system with \QANUS\ framework and additional components.}
	\label{fig:extensions}
\end{figure*}


The \QANUS\ framework adopts a pipelined approach to QA. The pipeline consists of four stages executed serially.. The stages include (1)~\emph{information source preparation}, (2)~\emph{question processing}, (3)~\emph{answer retrieval} and (4)~\emph{evaluation}. Within the framework we have implemented much of the programming code that will otherwise have been repeated across different QA systems. The framework can thus be likened to a foundation on top of which components can be added to obtain a complete QA system. 

Figure~\ref{fig:extensions} illustrates a complete QA system built with the framework. The upper-half of the figure delineates clearly the key classes that constitute the four stages of the framework listed earlier. The bottom-half of the figure shows additional components that can be added to the framework to complete the QA system. For completeness, the input and output to the various stages of the system are also depicted as shaded boxes at the bottom of the figure.

The top half of Figure~\ref{fig:extensions} shows that each of the stages share a common architecture, composed of two main classes.  The \texttt{FrameworkController} is responsible for directing the program flow and managing any input and output required or produced by the stage. It also invokes appropriate methods in the latter to process any input sent to the stage. The \texttt{FrameworkEngine} class provides the required processing that is needed on the various pieces of input to the stage. The processing that is required in each stage differs. For example, in the \emph{information source preparation} stage, processing may involve part-of-speech tagging an input corpus, while in \emph{question processing}, processing may instead be classifying the expected answer type of the posed questions. 

Due to space constraints, the individual interfaces and function calls presented by \QANUS\ are not explained in detail here. The full documentation together with the source code for the framework are available at the \QANUS\ download site\footnote{http://junbin.com/qanus}.

We briefly explain the operations that may be carried out in each stage. Note that this description serves merely as a guide, and users of the framework have full flexibility in deciding the operations to be carried out at each stage.



\vspace{2px}\indent\textbf{Information Source Preparation}. In this stage, an information source from which answers are to be obtained is set up. The framework is not restricted to any particular type of information source. Depending on the required needs and specifications, the eventual information source can be as varied as a \textsc{Lucene}\footnote{Open-source text search engine written in Java} index of the source documents, a full-fledged ontology or the Internet. Any necessary pre-processing to set up the information source is done here. 
Note that this stage prepares static information sources.  Using the Web dynamically as an information source is done in the subsequent answer retrieval stage.

\vspace{1px}\indent\textbf{Question Processing}. Typically, questions posed to the system need to be parsed and understood before answers can be found. Necessary question processing is carried out here. Typical operations here can include forming a query suitable for the information source from the posed questions, question classification to determine the expected answer type, as well as part-of-speech tagging and parsing. The outputs of these various operations are stored so that they can subsequently be used by the next stage in the QA pipeline.

\vspace{1px}\indent\textbf{Answer Retrieval}. The answer retrieval stage makes use of the annotations from the question  processing stage, and looks up the information source for suitable answers to the posed questions.  Incorporating candidate answers from dynamic sources, such as the Web or online databases, can also be incorporated here.   Proper answer strings that can answer the questions are extracted in this stage. If desired, answer validation can be performed as well.



\vspace{1px}\indent\textbf{Evaluation}. With the three stages above, \QANUS\ already provides the support necessary for a fully functional QA system. The \emph{evaluation} stage is introduced to complement the earlier stages and ease the verification of the performance of the developed QA system. It is optional and may be omitted if desired. The evaluation stage cross-checks the answers computed previously by the answer retrieval stage with a set of \textit{gold-standard} answers. The results of the evaluation are then output for easy review.  


\subsection{Additional Components}

The four stages of the \QANUS\ framework establish the flow of data through the entire QA pipeline, and form the backbone of any instantiated QA system. To realise the framework and obtain a fully functional QA system, additional components such as those shown in the bottom half of Figure~\ref{fig:extensions} must be coupled to the \QANUS\ framework.

The classes in the framework enforce the required interfaces that need to be adhered to by these additional components. By following the specified interfaces, any desired functionality can be plugged into the framework.


To give a better picture of how these components can be easily added to the \QANUS\ framework to complete a QA system, let us walk through an example for the \emph{question processing} (QP) stage. From Figure~\ref{fig:extensions}, the minimum set of components that need to be implemented for QP include the \texttt{QPController},  \texttt{QuestionInputHandler}, and \texttt{QPEngine}.


\vspace{2px}\indent\textbf{QPController}. \texttt{QPController} inherits from the \texttt{QPFrameworkController} component of the \QANUS\ framework. This component is responsible for initializing and integrating any text processing modules that will be used to process input questions with the framework. Suppose we want to perform part-of-speech tagging on the input questions, a part-of-speech component module needs to be created in \texttt{QPController}. \texttt{QPController} next notifies the \texttt{QPEngine} component about this part-of-speech tagger component.

\vspace{1px}\indent\textbf{QuestionInputHandler}. This component is responsible for reading in provided input questions. The implementation is thus dependent on how the input questions are formatted and presented. 

\vspace{1px}\indent\textbf{QPEngine}. This component is derived from the \texttt{QPFrameworkEngine} component of the \QANUS\ framework. It makes use of the earlier \texttt{QuestionInputHandler} component to read in input questions, and invokes any text processing modules registered with it by the \texttt{QPController} to annotate the question text. 

It is useful to emphasise here the ease and flexibility provided by the \QANUS\ framework: (1) The abstraction provided by the framework greatly reduces the amount of code that needs to be written for a QA system. 
Only a minimal set of customisation needs to be carried out to complete the implementation of the QP stage. (2) The framework is sufficiently flexible to allow for a range of QA systems to be built. In the explanation here, only a part-of-speech tagger is described. Depending on requirements, other text processing algorithms and techniques can also be incorporated.

\section{Implementation of \QANUSSYS }

To demonstrate the utility and practicality of the \QANUS\ framework, we have developed a QA system, referenced to as \QANUSSYS\ on top of the framework. The implementation of \QANUSSYS\ is included when downloading \QANUS\, to serve as an effective reference implementation and help reduce the learning curve for researchers in using the framework.

\QANUSSYS\ is a fully functioning QA system developed to run on the well-known dataset from the TREC 2007 QA track \cite{Dang2007}. \QANUSSYS\ makes use of IR-based techniques to perform the QA task. As can be seen later, this includes making use of a text search engine to perform document lookup, as well as lexicon-based techniques including named entity recognition for answer retrieval. An IR-based approach is adopted because it has been shown to turn in credible performances as explained earlier \cite{Hickl2007,Chali2007}.

\begin{figure*}[htbp]
	\centering
		\includegraphics[width=0.90\textwidth]{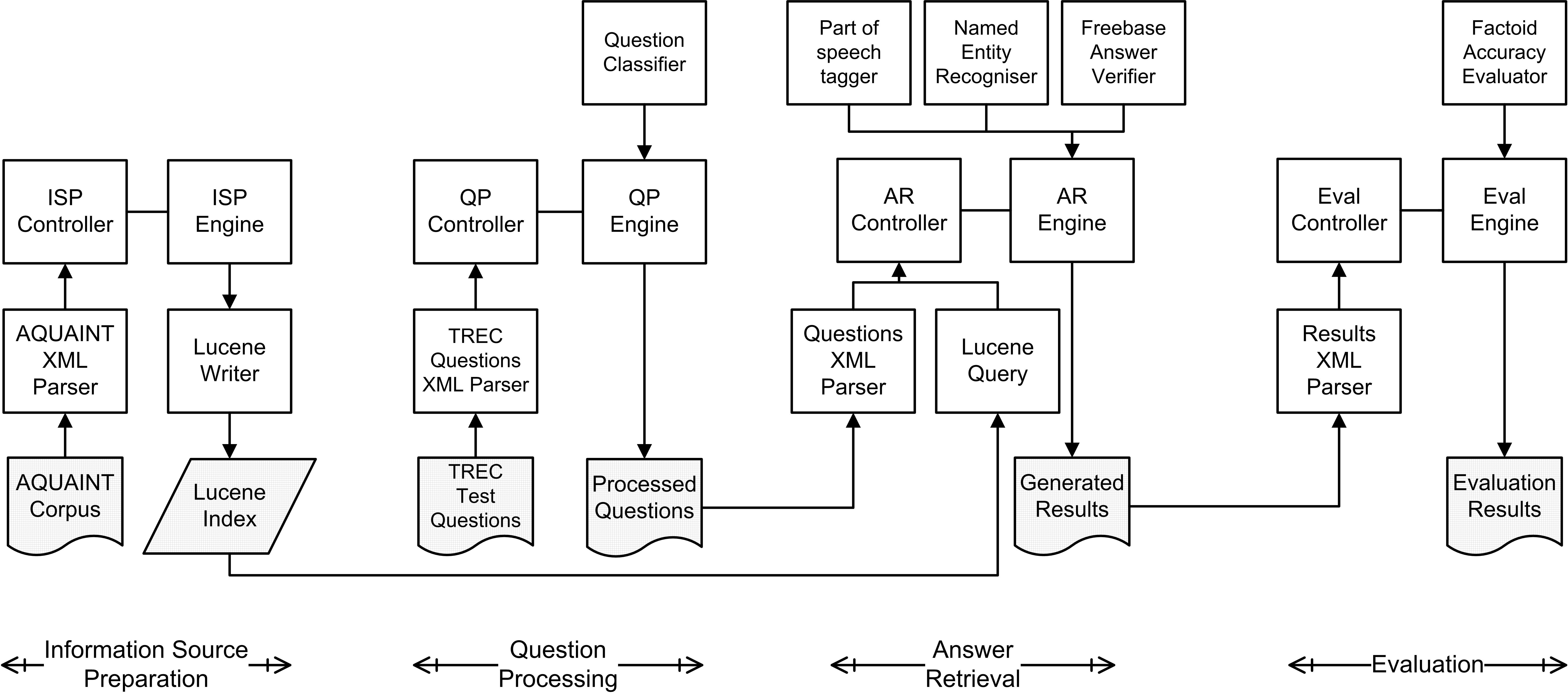}
	\caption{Actual components implemented in \QANUSSYS\ on top of the \QANUS\ framework.}
	\label{fig:actual-implementation}
\end{figure*}

Conforming to the description of the \QANUS\ framework, Figure~\ref{fig:actual-implementation} shows the various classes that have been implemented as part of \QANUSSYS . This figure is similar to Figure~\ref{fig:extensions}, which shows possible components needed to obtain a complete QA system. 

\vspace{2px}\indent\textbf{Information Source Preparation}. Similar to the participating machines of the TREC 2007 QA track, \QANUSSYS\ makes use of the \textsc{Aquaint-2} corpus\footnote{The corpus is not included with the download for \QANUSSYS\ as it is the intellectual property of the \textsc{Linguistic Data Consortium}.} which is stored in XML format. A XML parser \texttt{AQUAINTXMLParser} is written to interface the corpus with \QANUS . 
\texttt{LuceneWriter} makes use of \textsc{Lucene} to build an index of the input corpus. We will subsequently make use of this index to retrieve documents relevant to posed questions in the later stages of the QA pipeline.

\vspace{1px}\indent\textbf{Question Processing}. In this stage, \QANUSSYS\ attempts to classify the expected answer type of the input questions based on the taxonomy described in Li and Roth~\shortcite{Li2002} with \texttt{QuestionClassifier}. We built the classifier used by training the Stanford Classifier \cite{Manning2003} on the data described in Li and Roth~\shortcite{Li2002}. The classification assigned to each question is stored and passed on to the \emph{answer retrieval} stage.

\vspace{1px}\indent\textbf{Answer Retrieval}. To look up answers to the posed questions, \QANUSSYS\ form a query out of the question by dropping stop-words found in the question. \texttt{LuceneQuery} uses this query to search through the \textsc{Lucene} index built earlier in the \emph{information source preparation} stage. Documents retrieved by the \textsc{Lucene} search engine are then broken down into individual passages. \texttt{AnswerRetrieval} scores each of these passages using a variety of heuristics such as by tabulating the occurrences of the query terms within the passages. 

From the ranked passages, answer candidates are extracted depending on the expected answer type previously determined in  \emph{question processing}. For a question seeking a person name for example, a named entity recogniser \cite{Finkel2005} is used to extract candidate people names from the ranked passages. For other expected answer types such as dates, hand-written regular expressions are used to aid in the extraction of answer candidates.

%

Finally, the answer candidates are ranked based again on a set of heuristics which include the proximity of the candidates within the ranked passages to the query terms for example. The highest ranked candidate is returned as the preferred answer. 




\vspace{1px}\indent\textbf{Evaluation}. The evaluation stage provided by the \QANUS\ framework makes it possible to easily test the performance of \QANUSSYS . Currently \QANUSSYS\ supports only factoid questions, and so the evaluation metric used here is factoid \emph{accuracy} \cite{Dang2007}, defined as:
\begin{eqnarray*}
\textit{accuracy} = \frac{\textrm{no. of correctly answer questions}}{\textrm{total no. of test factoid questions}}
\end{eqnarray*}


\noindent which is implemented in \texttt{FactoidAccuracyEvaluator}. 

The top system in the TREC 2007 QA track \textsc{LymbaPA07} and the tenth-placed system \textsc{Quanta} achieved accuracy scores of 0.706 and 0.206 respectively. \QANUSSYS\ currently obtains an accuracy of 0.119.


There is room for improvement before \QANUSSYS\ can catch up with the state-of-the-art. 
The current implementation is simplistic and does not do much processing of the input questions, nor does it perform elaborate ranking of retrieved documents. As work on the system progresses and more sophisticated components are included into the system, \QANUSSYS\ should be able to achieve better results.

\section{Future Work}

\QANUS\ and \QANUSSYS\ are currently under development. \QANUS\ is relatively mature, having undergone several iterations of improvements and our work is now focused on improving the performance and functionalities of \QANUSSYS . 

\vspace{1px}\indent\textbf{Performance}. Conventionally, QA systems have been benchmarked against the systems participating in the TREC QA track. However recently the QA track has been dropped from both TREC and the Text Analysis Conference (TAC). As the years go by, the results from the QA track will age and become irrelevant.
There is also a trend towards the use of the Web as an aid for QA. The Web is dynamic and any such QA system will likely not generate the same results in different instances of time. For useful benchmarking, it is thus important to be able to use a baseline system which makes use of the Internet at the same time instance as the QA system being compared to. Having access to such a baseline system is thus critical and essential. This is the niche that \QANUSSYS\ serves to address.. When the performance of \QANUSSYS\ catches up with the state-of-the-art, it will be a useful baseline system against which other QA systems can be evaluated against.

To boost performance, more work needs to be done for the \emph{question processing} and \emph{answer retrieval} stages. There are plans to include a query expansion component which will be helpful in boosting the precision of the documents retrieved by \textsc{Lucene}. To improve on answer retrieval, \emph{soft} patterns as described in Cui {\it et al.}~\shortcite{Cui2007} can replace the current \emph{hard} hand-written patterns used in the system. More advanced measures like the use of dependency relations \cite{Cui2005} can also be adopted to improve on the current passage ranking implementation.

\vspace{1px}\indent\textbf{List questions}. Besides performance, it will also be useful to expand the functionalities of \QANUSSYS . It does not handle list questions for the moment. An implementation based on the use of redundancies found within the source text \cite{Banko2002,Lin2007} is being considered.

\vspace{1px}\indent\textbf{Internet front-end}. An online demonstration of \QANUSSYS\ is currently hosted online\footnote{http://wing.comp.nus.edu.sg/$\sim$junping/qanus/online/main.php} and supports querying over a pre-indexed \textsc{Aquaint-2} corpus or the Internet. The answer retrieval component working with data from the Internet is rudimentary and lacks techniques to process the noise that accompanies data downloaded from the Internet. It will be useful to improve on this Internet-querying component by adding better post-processing over the retrieved data.

\section{Conclusion}

The lack of community-available QA systems has made it difficult to create new QA systems and perform comparisons across published studies. This motivated our work on an open-source QA framework \QANUS . The framework implements much of the code needed for a QA system and reduces the development effort needed to build new systems. It is carefully designed to be flexible and supports the use of a wide range of QA techniques.

As a demonstration of the utility and practicality of \QANUS , we have also implemented a fully functional factoid QA system \QANUSSYS\ on top of the framework. Our goal is to improve \QANUSSYS\ so that it will serve as a useful and accessible baseline to benchmark future QA systems and technologies against.
Through this work, we hope to lower the high barriers of entry facing new QA researchers and reduce the time needed for them to begin productive research in this area.

%

\bibliographystyle{acl}
\bibliography{Bibliography}

\end{document}